\documentclass{PoS}
\def\degr{\hbox{$^\circ$}}
\def\fdg{\hbox{$.\!\!^\circ$}}
\def\farcs{\hbox{$.\!\!^{\prime\prime}$}}

\title{X-shaped radio sources as parent population\\
of core-dominated triple blazars}

\ShortTitle{X-shaped radio sources as parent population of core-dominated triple blazars}

\author{\speaker{Andrzej Marecki}\\
        Toru\'n Centre for Astronomy, N. Copernicus University, Toru\'n, Poland\\
        E-mail: \email{amr@astro.uni.torun.pl}}

\abstract{There are a number of theories explaining the nature of the 
so-called X-shaped radio sources. According to one of them, an X-shaped 
source is indeed a cross whose one arm is associated with a double radio 
source that has changed its orientation in space, while the other arm is 
associated with relic lobes and its position indicates the former direction 
of the jets. Here, I present two new arguments in favour of this conjecture. 
Firstly, it is obvious that shortly after the repositioning, the pair of the 
new lobes must be very compact. To illustrate such a possibility, I show an 
EVN\footnote{The EVN is a joint facility of European, Chinese, South African
and other radio astronomy institutes funded by their national research
councils.}
image of the central component of a triple source J1625+2712. When 
resolved, it appears as a compact double that is not aligned with the outer 
double so the whole source is indeed X-shaped. Secondly, I consider the 
situation when one of the arms of an X-shaped source is not intrinsically 
short but foreshortened by projection. I show two examples of triple sources 
whose central component is a blazar and the span of the lobes that straddle 
it amounts to more than $6\times10^5$\,pc. An assumption that sources of 
this kind have one axis, and so the lobes are beamed in the same way the 
core is, would require unrealistically huge deprojected linear sizes. 
Therefore, I claim that core-dominated triples (CDT) like these two have two 
axes: the one pertinent to the relic lobes is not pointed towards us so they 
are not beamed/foreshortened, whereas the axis pertinent to the jets makes a 
small angle with the line of sight so that a blazar is observed. It follows 
that X-shaped sources must be actual crosses and they are the parent 
(unbeamed) population of at least some CDT blazars, particularly those with 
large overall sizes.}

\FullConference{Resolving The Sky -- Radio Interferometry: Past, Present and Future \\
		 April 17-20,2012\\
		 Manchester, UK}

\begin{document}

\section{Introduction}

For decades, the ``twin exhaust'' model of a classical double radio source 
\cite{BR74} along with the unification scheme in which the orientation 
effects play a fundamental role \cite{UP1995} have been used as a general 
theoretical framework successfully explaining the nature of an overwhelming 
majority of radio sources identified with active galactic nuclei (AGN). 
Nevertheless, it has been noticed \cite{LW1984} that some extragalactic 
radio sources can have more complicated structures. Apart from the 
components accounted for by the theory -- the cores, the jets, and the lobes 
-- ``bridges'' linking the host galaxy with the outer edges of a radio 
source are also evident. In \cite{LW1984}, bridges are interpreted as a 
result of backflows, i.e. two streams of matter flowing back from the 
hotspots towards the host galaxy. Moreover, it has been claimed in 
\cite{LW1984} that when the backflowing material originating at one hotspot 
collides with the backflow from the opposite lobe, it may expand laterally 
giving raise to a thick disk located perpendicularly to the lobe axis. When 
viewed from the side, such a disk would be perceived as a secondary 
elongated structure, that, together with the primary one consisting of the 
lobes and bridges, would look as two arms of a cross. Three objects 
described there -- 3C\,52, 3C\,136.1, and 3C\,315 -- clearly bore the 
signatures of cross-shaped (hereafter X-shaped) radio sources. It must be 
stressed though that in the backflow scenario, these sources are not crosses 
intrinsically but only apparently, hence the ``X-shape effect'' takes place 
only when the axis of the source lies in the sky plane and so the disk 
resulting from the clash of backflows is seen from the side.

An opposite concept that X-shaped sources {\em are} actual crosses is 
plausible but requires a completely different mechanism -- the jet 
re-orientation. When it happens, the existing standard double-lobed 
structure is no longer fuelled so it becomes weak and diffuse, and its 
spectrum steepens owing to radiation and expansion losses, while a new 
double-lobed structure centred at the same host galaxy -- it is oriented at 
an angle with regard to the other one -- starts to build up. As a result, an 
X-shaped structure develops. The primary arm normally resembles 
Fanaroff--Riley type~II (FR\,II) \cite{FR74} radio source, while the 
secondary arm -- the two halves of which are often termed ``wings'' -- is a 
relic whose position indicates the former orientation of the jet. A 
comprehensive analysis of the properties of radio emission from several best 
known X-shaped sources presented in \cite{helgephd} lends a strong support 
to this interpretation.

The strengths and weaknesses of different models of X-shaped sources were 
critically assessed in \cite{LR2007}. Based on the two-frequency 
observations of eleven X-shaped sources carried out with the Giant Metrewave 
Radio Telescope (GMRT), the authors concluded that backflow as well as 
buoyancy and conical precession did not convincingly explain formation of 
X-shaped sources, thereby re-orientation of the jet axis owing to a merger 
appears as a plausible solution. Yet, at least one issue can hardly be 
reconciled with this particular interpretation -- the small number of known 
X-shaped sources as compared to the ubiquity of mergers. That's why the 
authors of \cite{LR2007} proposed their own concept, in which the scarcity 
of X-shaped sources stems from a rare putative situation when a galaxy 
contains two active nuclei each responsible for a pair of jets. However, 
there is a serious problem with this approach -- it fails to explain the 
absence of X-shaped sources where both lobe pairs are of FR\,II type -- see 
e.g. Fig.\,1 in \cite{Cheung2007} -- which should be quite a likely 
combination if we assume the presence of two independent central engines. It 
looks, therefore, that also this model is not satisfactory and, 
consequently, neither one out of the several available in the literature to 
date provides a fully consistent explanation of the nature of X-shaped 
sources.

Does this mean that the correct theory of X-shaped sources is yet to be 
found? In the light of \cite{Mezcua2011}, probably not. Assuming that one of 
the existing models -- a sudden flip of the jet orientation caused by a 
merger -- is the most promising one, the authors put it under a rigorous 
test. Its idea is as follows. If mergers are the prime cause that eventually 
leads to creation of X-shaped sources then the supermassive black holes 
(SMBH) in the centres of their host galaxies should statistically have 
higher masses than central SMBHs in standard double-radio-lobed active 
galaxies. Also, a past merger event should be reflected in starburst history 
of the host galaxy. In order to test these conjectures, the SMBH masses, 
luminosities, starburst histories, and jet dynamic ages in a sample of 29 
X-shaped galaxies selected from the list of candidates extracted from the 
VLA FIRST\footnote{{\em Faint Images of the Radio Sky at Twenty 
(Centimetres)}} catalogue \cite{Cheung2007} have been determined and 
compared with those in a control sample of 36 radio-loud active galaxies 
with similar redshifts and optical and radio luminosities \cite{Mezcua2011}.

The outcome of the test provided a strong underpinning to the hypothesis of 
jet re-orientation induced by a merger. The mean SMBH mass of X-shaped 
sources' host galaxies indeed appeared higher than that for the control 
sample as predicted by the model. A comparison of the starburst and dynamic 
ages in the galaxies hosting X-shaped sources and those in the control 
sample lent further support to the scenario in which the primary arm of an 
X-shaped source might have resulted from re-orientation of the host galaxy's 
SMBH \cite{ME2002} following the parent pair of black holes coalescence, the 
ultimate stage of a merger. In particular, it was showed in \cite{ME2002} 
that the orientation of a black hole spin axis would change dramatically 
even in a minor merger, leading to re-orientation of the coalescing SMBHs 
and a flip in the direction of the jets. On a basis of a thorough 
theoretical analysis, Liu \cite{Liu2004} concluded that the realignment of a 
rotating SMBH with a misaligned accretion disk was due to the 
Bardeen-Petterson effect \cite{BP1975} and that the timescale of such a 
realignment was $<10^5$\,years, which was negligible in comparison with 
typical lifetime of the lobes that are no longer fuelled (up to 
10$^8$\,years \cite{KG1994}). These two estimates fit well to the likely 
mechanism of creation of X-shaped objects. On the one hand, the realignment 
process is relatively quick so the new position of the jets is established 
early allowing the new (primary) arm to grow for sufficiently long time to 
attain a noticeable size, which is what we observe. On the other hand, owing 
to its longevity, the old (secondary) arm remains visible during all that 
process.

The weight of observational evidence after publication of \cite{Mezcua2011} 
seems to favour the scenario in which merger is the prime cause triggering 
creation of an X-shaped source. It must be noted, however, that it has one 
weakness: X-shaped sources are rarely identified with mergers, albeit an 
impressive example of an X-shaped source associated with a merger was 
recently shown in \cite{KW2012}.

\section{In quest of concealed X-shaped sources}

There is yet another important prediction of the central engine 
re-orientation scenario: the existence of ``concealed'' X-shaped sources. 
When the axis of the central engine changes its position (whatever the 
mechanism responsible for this -- merger or maybe something else), the new 
primary arm starts ``from scratch''. Therefore, if only the re-orientation 
is a real phenomenon then there should exist a rare subclass of sources 
where an event of this kind took place quite recently so that the length of 
the newly established primary arm is so far minute compared to the length of 
the old, large-scale secondary arm. In practical terms, this means that a 
newly created X-shaped source would {\em not} be perceived as such because 
the very short primary arm would not be resolved in an image encompassing 
the whole source. For example, if repositioning of the central engine took 
place, say, $10^5$\,years ago then a new double growing at a rate of 
$0.2-0.3c$ would attain the length of the order of only a kiloparsec. It 
follows that the primary arm will be embedded within a source compact enough 
to appear in a map of the whole radio structure as a single, point-like 
component straddled by large-scale double.

In 2004, I launched an observational project aimed at discovering of a 
number of exotic classes of radio sources including the one characterised 
above. The idea was as follows. Clearly, there exist radio sources somewhat 
similar to FR\,II, where the FR\,II-like lobes straddle a central component 
which is by far the brightest feature of the whole object. I labelled them 
core-dominated triple (CDT) sources. CDTs are intriguing because true 
FR\,IIs are lobe-dominated so there must be a fundamental difference between 
these two classes. There are three possible explanations of CDTs:

\begin{enumerate}

\item The ``core'' of a CDT is actually a compact double aligned with the 
large-scale (outer) double. As a whole, the object is a double-double radio 
source with an extreme ratio of outer/inner double linear sizes.

\item The ``core'' of a CDT is actually a compact double but not aligned 
with the large-scale double. This is a ``concealed'' X-shaped source shortly 
after repositioning of the central engine.

\item The core of a CDT is actually a radio-loud AGN like the core of a 
standard FR\,II source but its excessive brightness is a result of Doppler 
boosting caused by beaming. As a whole, the object is in fact a fully 
fledged X-shaped source where the primary arm is pointed towards the 
observer and so extremely foreshortened, while the secondary is not.

\end{enumerate}

More than a hundred CDTs were selected from FIRST using a semi-automated 
procedure described in \cite{Mar2006}. I divided the sources into two groups 
depending on whether the ``cores'' had steep or flat spectrum. Objects in 
the steep-spectrum group could possibly belong to the first or the second 
category out of the there itemised above, whereas flat-spectrum objects are 
likely to belong to the third category. The central component of the members of 
the steep-spectrum group were observed with MERLIN \cite{Mar2006} and then, 
depending on their nature revealed by MERLIN, possibly followed up with the 
EVN. This observational campaign already brought one interesting result; it 
was shown in \cite{MS2009} that the first category of object has at least 
one specimen: B\,0818+214. It was resolved with MERLIN at 6\,cm and the 
EVN observations at 18\,cm confirmed it was a double, not a core-jet. All in 
all, B\,0818+214 turned out to be an extreme case of a double-double source, 
where the linear size of the inner double was two orders of magnitude less 
than the size of the outer double. Six sources observed with MERLIN had 
unresolved cores, though -- see Table\,1 in \cite{Mar2006}. Thus, a 6-cm EVN 
follow-up was carried out to resolve them. In the case of J1625+2712, one of 
those six targets, the EVN result was very surprising.

J1625+2712 is a CDT source where, according to FIRST, the core of 261\,mJy 
is straddled by weak lobes of 5 and 1\,mJy, respectively. The spectral index 
of the core between 1.4\,GHz and 5\,GHz is $\alpha=-0.68$ so it is clearly a 
steep spectrum source. The map in the left panel of Fig.\,1 is a cut-out 
from FIRST showing the whole J1625+2712. Not only in this image but also in 
MERLIN 6-cm map, the core is point-like. However, in the 6-cm EVN map, the 
alleged ``core'' has been resolved into a double where the separation of the 
two peaks is only 18 milliarcseconds -- see the right panel of Fig.\,1. The 
P.A. of the line connecting the maxima is $-7\fdg6$, whereas the P.A. of the 
large-scale structure -- its angular size reaches $53\farcs5$ -- is 
$42\degr$. It follows that the misalignment between the two pairs of lobes 
is nearly $50\degr$.

\begin{figure}[ht]
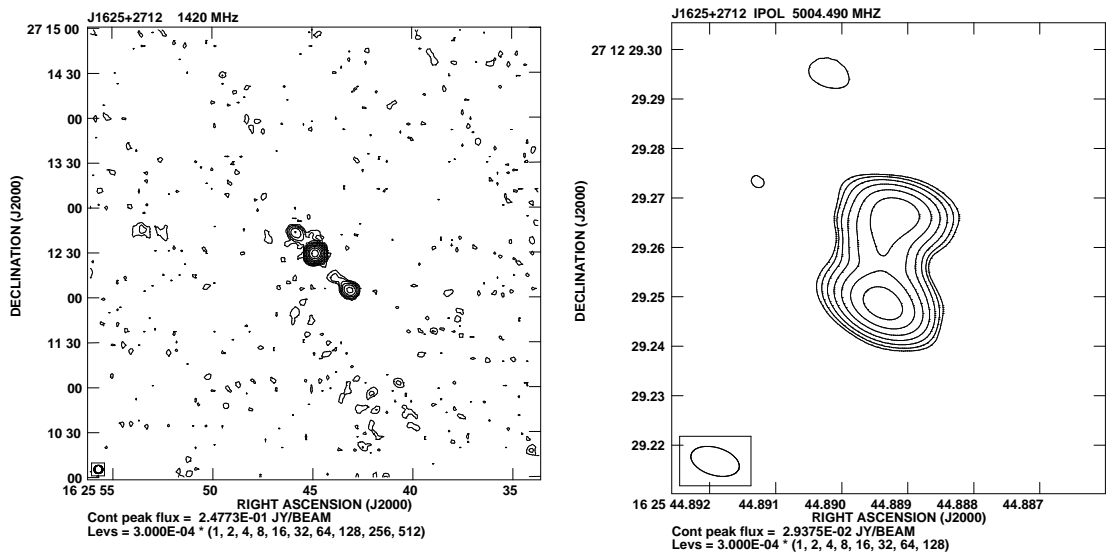

\centering
\includegraphics[width=.49\textwidth]{J162544+271229.ps}
\includegraphics[width=.49\textwidth]{J1625+2712.EVN.ps}
\caption{{\em Left}: A cut-out from FIRST centred on J1625+2712,
{\em Right}: a 6-cm EVN image of the central component seen in the FIRST map}
\end{figure}

The most straightforward interpretation of J1625+2712 is that this is an 
X-shaped source, but since one arm of the cross is 3,000 times shorter than 
the other, the actual structure of J1625+2712 cannot be recognised in the 
FIRST map showing the whole radio source. Unfortunately, the redshift of 
J1625+2712 is not known, so we can only estimate the upper limit for the 
linear size\footnote{Based on the well-known property of the angular-size
distance: the ratio of linear to angular sizes cannot exceed
8.558\,pc/milliarcsecond (assuming standard cosmological parameters).}
of the compact double, which is 154\,pc
(500\,ly). Thus, I claim that J1625+2712 is a ``concealed'' X-shaped 
source observed very shortly after the repositioning of the central engine. 
The case of J1625+2712 supports the hypothesis that X-shaped sources are 
actual crosses.

\begin{figure}[t]
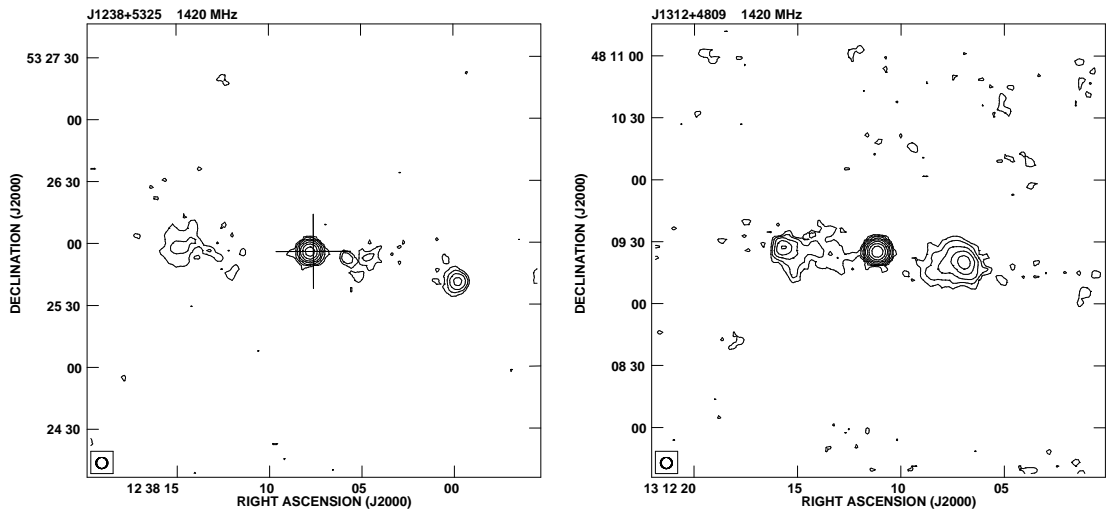

\centering
\includegraphics[width=0.49\columnwidth]{J1238+5325.ps} 
\includegraphics[width=0.49\columnwidth]{J1312+4809.ps}
\caption{FIRST images of two CDT blazars. Contours are increased by a 
factor of 2; the first contour level corresponds to 0.4\,mJy/beam.}
\end{figure}

\section{Distortion of X-shaped sources caused by orientation}

Because of the growing weight of evidence that X-shaped sources {\em are} 
actual crosses, it becomes more and more obvious that we should consider the 
impact of orientation on the perceived appearance of such sources. Plain 
X-shaped structure requires absence of any significant distortion caused by 
orientation effects, i.e. both arms of the cross must lie close to the sky 
plane for an X-shaped source to be observed as such. Let's assume that in 
some X-shaped sources only the secondary arm lies close to the sky plane 
while the primary arm is oriented towards the observer. In this 
configuration, the observed structure would consist of a beamed, 
Doppler-boosted base of the jet seen head-on and so perceived as a strong 
single component (a core), whereas the wings would appear as a diffuse 
double straddling that bright core. We would thus observe a CDT source with 
a flat-spectrum central component. When the angle between the primary arm 
and the line of sight is small, then, according to the unification scheme 
of AGNs \cite{UP1995}, the core component would be perceived as a blazar.

If the projected span of the lobes in a CDT identified with a blazar is of 
the order of a significant fraction of a megaparsec, an extreme 
foreshortening of the whole source imposed by the presence of blazar in its 
centre is rather impossible, otherwise the actual size of the source would 
be prohibitively large. However, it is very easy to circumvent this problem 
-- the foreshortening of the outer components of a CDT blazar does not need 
to be large or may even not take place at all if we assume that they 
correspond to the secondary arm of an X-shaped source and so they are not 
beamed. I thus hypothesise that X-shaped sources are the parent population 
of at least some CDT blazars, especially those whose overall projected 
linear sizes of the triple are large.

Fig.\,2 presents two blazars with obvious CDT large-scale structures. Their 
basic parameters are shown in Table\,1. As can be seen there, the overall 
linear sizes of the radio structures are of the order of hundreds of 
kiloparsecs. If one still requires that these sources are standard FR\,II 
doubles but highly foreshortened so that their cores appear as blazars 
\cite{AU1985}, then the deprojected linear sizes would be at least of an 
order of magnitude higher. The linear size of the largest radio source known 
to date -- a giant radio galaxy (GRG) J1420$-$0545 -- amounts to 4.69\,Mpc 
\cite{Mach2008}. Although it cannot be ruled out completely that the two 
objects shown in Fig.\,2 are in fact extremely large but highly 
foreshortened GRGs, my claim that the outer components of CDTs are not 
subject of extreme foreshortening appears as a competitive alternative.

\begin{table}[t]
\caption{Basic parameters of CDT blazars that are likely to be X-shaped sources
distorted by orientation}
\begin{center}
\begin{tabular}{c c c c r r r c}
\hline
\hline
Name & RA & Dec & $z$ & $S_{core}$ & $S_{lobe1}$ & $S_{lobe2}$ & Linear size\\
\hline
J1238+5325 & 12 38 07.782 & +53 25 55.83 & 0.347 &  43.24 &  8 &  7 & 663 \\
J1312+4809 & 13 12 11.144 & +48 09 25.22 & 0.715 & 105.70 & 36 & 21 & 625 \\
\hline
\end{tabular}
\end{center}
\small
Coordinates and $S_{core}$ are from FIRST catalogue. $S_{lobe1/2}$ 
have been extracted from FIRST maps using AIPS utility TVSTAT. Fluxes (at 
21\,cm) are in mJy. Linear sizes are in kpc. Standard values for $H_0$, 
$\Omega_m$, and $\Omega_{\Lambda}$ have been used to calculate the linear 
sizes.
\end{table}

\eject
The above proposal has several advantages:

\begin{enumerate}

\item The overall projected linear sizes of CDT blazars can be even of the 
order of a megaparsec simply because there is little or hardly any 
projection acting on the large-scale structure.

\item The reason why the outer components of a CDT blazar are diffuse is 
obvious: they look so because they are fading since they are no longer 
fuelled by the jets that are now oriented towards a completely different 
direction: close to the line of sight.

\item The presence of misalignments between the apparent direction of the 
jet in the central component of a CDT (if resolved) and that of the 
large-scale structure -- such a phenomenon is clearly observed in blazars in 
general \cite{Appl1996,Kharb2010} -- is obvious, expected and even required. 
This is because the direction of the primary arm is close to the line of 
sight as posited by the blazar paradigm, but usually not identical. 
Therefore, apart from having a dominating radial component, this direction 
has also a minor tangential component. The observed positional angle of the 
latter, i.e. the apparent orientation of the jet on the sky plane, is random 
and so uncorrelated with any fixed orientation including that of the 
secondary arm.

\end{enumerate}

\section{Conclusions}

My preferred interpretation of the nature X-shaped sources is that they are 
just standard doubles that have undergone repositioning of the axis. The 
``wings'' are relics of the former lobes. I also propose that CDTs are 
closely related to X-shaped sources. Two scenarios are possible: CDTs with 
steep-spectrum central component are those X-shaped sources that are seen 
shortly after the repositioning and so the newly created pair of lobes is 
still very compact, whereas CDTs with flat spectrum central component are 
typical X-shaped sources where the primary arm of the cross is beamed 
towards the observer. The latter scenario is particularly likely for CDTs 
whose central components are identified with blazars.

If the above two-part interpretation of CDTs is correct then X-shaped 
sources are perceived as cross-shaped simply because they {\em are} actual 
crosses, not apparent owing to a particular orientation as posited by the 
backflow model. Hence, the explanation presented here contradicts the 
backflow model and strongly underpins the AGN re-orientation scenario 
proposed in \cite{ME2002} and later elaborated in \cite{Liu2004}.

\end{document}